\documentclass[]{mn2e}  
\usepackage{times}

\def\heao1{{\it HEAO-1\/}}

\newcommand{\ltsima}{$\; \buildrel < \over \sim \;$}
\newcommand{\simlt}{\lower.5ex\hbox{\ltsima}}
\newcommand{\gtsima}{$\; \buildrel > \over \sim \;$}
\newcommand{\simgt}{\lower.5ex\hbox{\gtsima}}

\def\lesssim{\mathrel{\hbox{\rlap{\hbox{\lower4pt\hbox{$\sim$}}}\hbox{$<$}}}}
\def\gtrsim{\mathrel{\hbox{\rlap{\hbox{\lower4pt\hbox{$\sim$}}}\hbox{$>$}}}}

\usepackage{graphicx}
\usepackage{txfonts}
\usepackage{rotating}

\title{The ages of stellar populations in a warm dark matter universe} 
  \author[F. Calura et al.]  {F. Calura$^{1}$\thanks{E-mail:
      fcalura@oabo.inaf.it}, N. Menci$^{2}$, A. Gallazzi$^{3,4}$\\
(1) INAF-Osservatorio Astronomico 
di Bologna, Via Ranzani 1, 40127 Bologna, Italy\\ 
(2) INAF-Osservatorio Astronomico di Roma, via di Frascati 33, 00040 Monte Porzio Catone, Italy\\
(3) INAF-Osservatorio Astrofisico di Arcetri, Largo Enrico Fermi 5, I-50125 Firenze, Italy;\\ 
(4) Dark Cosmology Center, University of Copenhagen, Niels Bohr Institute, 
Juliane Maries Vej 30, 2100 Copenhagen, Denmark\\}

\pagerange{\pageref{firstpage}--\pageref{lastpage}}
\pubyear{---}

\begin{document}

\maketitle

\label{firstpage}
\begin{abstract}
By means of a semi-analytic model 
of galaxy formation, we show how the local observed relation between age and galactic stellar mass 
is affected by assuming a DM power spectrum with a small-scale cutoff.  We compare results obtained by means of both a $\Lambda$-cold dark matter ($\Lambda$CDM) and a $\Lambda$-warm dark matter ($\Lambda$WDM) power spectrum - suppressed with respect to the $\Lambda$CDM at scales below $\sim 1$ Mpc.
We show that, within a LWDM cosmology with a thermal relic particle mass of 0.75 keV, both the mass-weighted and the 
luminosity-weighted age-mass relations are steeper than those obtained within a $\Lambda$CDM universe, in better agreement with 
the observed relations. Moreover, both the observed differential and cumulative age distributions 
are better reproduced 
within a $\Lambda$WDM cosmology.  
In such a scenario, star formation appears globally delayed with respect to the $\Lambda$CDM, in particular 
in low-mass galaxies. The difficulty of obtaining a full agreement between model results and 
observations is to be ascribed to our present poor understanding of baryonic physics. 
\end{abstract} 


\section{Introduction} 
The current `concordance' cosmological paradigm, which envisages a Universe 
dominated by cold dark matter (DM) and a dark energy $\Lambda$, 
predicts a `bottom-up' formation of structures, in which the smallest 
DM halos are the first to collapse, then interact and later merge to form larger halos. 
This $\Lambda$-cold dark matter ($\Lambda$CDM ) 
scenario is able to account for several lines of evidence, which concern primarily 
the large-scale properties of the Universe  and which 
include the temperature anisotropies in the microwave background radiation (Komatsu et al. 2011), galaxy clustering (Percival et al. 2010)  
and the observed flux power spectrum of the Lyman-$\alpha$ forest (Viel et al. 2005). 

This picture motivated several theoretical investigations of the properties of galaxies, 
which soon evidenced some major problems in the dominant cosmological picture, in particular 
on galactic and sub-galactic scales, i.e  below a few Mpc.   
First of all, N-body simulations have shown that a $\Lambda$CDM  cosmology implies  
inner density profiles much steeper than those inferred from the rotation curves of real galaxies (Moore et al. 1999, Gentile et al. 2007). 
In addition, $\Lambda$CDM-based simulations overpredict the number of satellites 
residing in  Milky Way-sized halos by $5-10$ times (Kauffmann et al. 1993; Moore et al. 1999, Klypin et al. 1999), a discrepancy 
commonly referred to as the Missing Satellite problem. 

It is worth stressing that other explanations could be found for the missing satellite problem. 
On the observational side, this problem could be due to the inability 
of modern telescopes to detect the faintest galaxies (Bullock et
al. 2010), 
or to an overestimate
of the virial mass of the Milky-Way dark matter halo (Wang et al. 2012). 
Moreover, it was shown that various baryonic processes, as well as large-scale effects such as `cosmic web stripping' (Benitez-Lambay et al. 2013) 
can play some role in reducing the abundance of low-mass objects in MW-sized halos (Sawala et al. 2013).  
However, it was also shown that the tension between the dynamical properties of Milky-Way satellites and 
$\Lambda$CDM expectations are more difficult to reconcile by means of baryonic processes 
(Boylan-Kolchin, Bullock \& Kaplinghat 2011; 2012).

Other problems related to the observable properties of stellar populations 
have been evidenced by $\Lambda$CDM-based semi-analytical galaxy formation models, which 
allow one to link directly baryon-related physical processes such as star formation, cooling and feedback (Cole et al. 2000) 
to the merging histories of DM haloes. 
In the past decades, semi-analytical models (SAMs) have turned out as useful tools to obtain a physically-motivated statistical view of the global properties of galaxies 
(Baugh 2006; Silk \& Mamon 2012). 
SAMs are commonly used to investigate the main galactic observables related to 
their star formation histories, their stellar mass and halo mass assembly, including 
the evolution of the luminosity function (e. g., Somerville et al. 2012, Menci et al. 2012), 
the cosmic star formation history (e. g., Nagashima et al. 2004), as well as the 
basic scaling relations (e. g., Fontanot et al. 2009) and the evolution of their chemical abundances (e. g., Calura \& Menci 2009; Yates et al. 2013),
often outlining further problems in reproducing the observations within $\Lambda$CDM models, again mostly related 
to the small-scale power excess which manifests in an excess of of faint systems at any redshift. 

Although it is still possible that the failure of an accurate description of the statistical  
properties of galaxies may be due to 
an incomplete theoretical understanding of the basic 
physics regulating the most fundamental baryonic processes (see, e.g., Koposov et al. 2009, Kang et al. 2013, Brooks et al. 2013), 
the above problems could also be related to the intimate nature of dark matter. 
In fact, the intrinsic excess of substructure typical of the $\Lambda$CDM scenario is less pronounced in 
a Universe dominated by Warm Dark Matter (WDM).
Compared to CDM, the WDM constituents are lighter particles with masses of a few keV,  
and hence with 
larger thermal speeds.
In terms of power spectrum, this corresponds to a suppression of the growth of
structures on the smallest scales, hence providing a natural mechanism to limit the 
wealth of substructures typical of the CDM scenario. 

The effects of a WDM power spectrum on the statistical properties of galaxies have been investigated 
in Menci et al. (2012), where it was shown that the adoption of a WDM scenario helps alleviate the 
tension between model predictions and the most small-scale sensitive observables, such as the faint end of the luminosity function. 
Also, the evolution of the active galactic nucleus (AGN) luminosity function (LF),  suffering the same problems as the galactic LF, appears better 
described with the assumption of an underlying WDM power spectrum (Menci et al. 2013). 

So far, little attention has been paid to another long-standing problem 
related to the properties of baryons, i.e. the inability of $\Lambda$CDM models to account for the 
observed relation between the luminosity-weighted age of the stellar populations and 
the galactic stellar mass. 
The observational study by Gallazzi et al. (2005) indicates that, on average, 
the larger the galaxy, the older its stellar populations, 
supporting the well-established galactic `downsizing' (Cowie et al. 1996; Mateus et al. 2006; Fontanot et al. 2009). 
In principle, this result is not inconsistent with the CDM scenario; in fact, the progenitor of massive galaxies are predicted to form 
in biased, high-density regions of the density field where star formation starts at early epochs. At later cosmological epochs, 
bottom-up hierarchical structure formation then assembles these early-formed stellar populations into massive, low-redshift galaxies.  
However, $\Lambda$CDM  models also predict the early collapse of a huge number of low-mass halos, which remain  isolated at later times 
retaining the early-formed stellar populations; as a result, $\Lambda$CDM-based SAMs  generally provide flat age-mass relations (Fontanot et al. 2009; Pasquali et al. 2010; De Lucia \& Borgani 2012), indicating a basic coevality of satellites and central galaxies. 
The suppression of the small-scale power due to a different 
free streaming length is expected to have effects not only on the 
abundance of collapsed objects below a cut-off scale, but also 
on the time of formation of halos (Angulo et al. 2013), consequently 
of the formation and assembly history of baryonic structures and ultimately on 
the ages of stellar populations. 
In this paper, our aim is to study 
the effects of a WDM cosmology on the star formation of galaxies. 
In particular, we will 
address the issue of the ages of stellar populations 
in galaxies of various masses within a WDM-based semi-analytical model (SAM) for galaxy formation, 
and we will test whether such a scenario can help alleviate the above discrepancy. \\
This paper is organized as follows. In Sect. 2, we briefly describe the main 
ingredients of the SAM used in this work. Our results are presented in Sect. 3. 
Finally, our conclusions are drawn in Sect. 4.

\section{Model Description} 
The effects of the WDM and CDM  power spectrum on the properties 
of stellar populations are investigated by means of 
the Rome semi-analytic model (R-SAM). 
This model connects 
the physical processes involving baryons (physics of the gas,
star formation, feedback) to
the merging histories of DM halos, determined by
the DM initial power spectrum. 
The main free parameters of the model have been described 
in Menci et al. (2005, 2006, 2008). 
Here, we recall the basic features of the model, the importance of the cosmological background 
in determining the properties of baryonic structures and 
a basic description of the main baryonic physics.

\subsection{Power spectrum and history of dark matter halos}
DM halos originate from the gravitational instability of tiny perturbations 
in the primordial density field, 
represented by a random, Gaussian density field.  
We assume a “concordance cosmology” (Spergel et al. 2007), characterised 
by a flat Universe and a matter density parameter $\Omega_M=0.3$ (with  $\Omega_b=0.04$ for 
the baryonic matter density parameter), and 
an Hubble constant (in units of
100 km s$^{-1}$ Mpc$^{-1}$)  $h=0.7$. 

As cosmic time increases, larger and larger regions of the density field collapse; 
in this picture, groups and clusters of galaxies grow by merging with mass- and redshift- 
dependent 
rates according to the Extended Press \& Schechter (EPS) formalism (Press \& Schechter 1974). 
The merging histories and the mass distribution of DM halos are determined by the 
linear power spectrum $P(k)$ computed for a sphere of radius $r$ 
at the wavelength $k=2\pi/r$ and defined as 
\begin{equation}
P(k)\equiv \left| {\delta ^2(\vec {k})}\right|, 
\end{equation}

where $\delta (\vec {k})$ is the density contrast in the 3-dimensional Fourier space (Einasto 2001; Norman 2001). 
For the CDM cosmology, we adopt the power spectrum $P_{CDM}(k)$ as given by Bardeen et al. (1986). 

In the case of WDM cosmology, the power spectrum is suppressed with respect to the CDM 
below a characteristic scale which  
depends on the mass of the WDM particles 
(for non-thermal particles, also on their mode of production; Kusenko 2009); 
the lighter the WDM particles and the larger their thermal velocities, the 
larger the free-streaming length $r_{fs}$, , i.e. the scale below which the perturbations are wiped out. 
If WDM consists of relic thermalized particles, the suppression in the power spectrum  $P_{WDM}(k)$, following 
the parametrization of Bode, Ostriker \& Turok (2001) and  Viel et al. (2005), 
can be quantified through the ratio of the two linear power spectra:

\begin{equation}
{P_{WDM}(k)\over P_{CDM}(k)}=\Big[1+(\alpha\,k)^{2\,\mu}\Big]^{-5\,\mu}\\
\end{equation}
with 
\begin{equation}
\alpha=0.049 \, \Big[ {\Omega_X\over 0.25}\Big]^{0.11}\, \Big[{m_X\over {\rm keV}}\Big]^{-1.11}\,
\Big[{h\over 0.7}\Big]^{1.22}\,h^{-1}\,{\rm Mpc}
\end{equation}
and $\mu=1.12$.
A similar relation holds if the WDM components are  sterile neutrinos, provided one substitutes the mass $m_X$ with a mass \\$m_{sterile}=4.43\,{\rm KeV}\,
(m_X/{\rm keV})^{4/3}\,(\Omega_{WDM}\,h^2/0.1225)^{-1/3}$ . 
In both cases, the smaller the WDM mass $m_X$ (or $m_{sterile}$), the larger the suppression with respect to the standard CDM spectrum. 
As a reference case, and for continuity with our previous works here we adopt a WDM thermal particle mass $m_X=0.75$ keV.

\begin{figure*} 
\includegraphics[width=130mm,height=90mm]{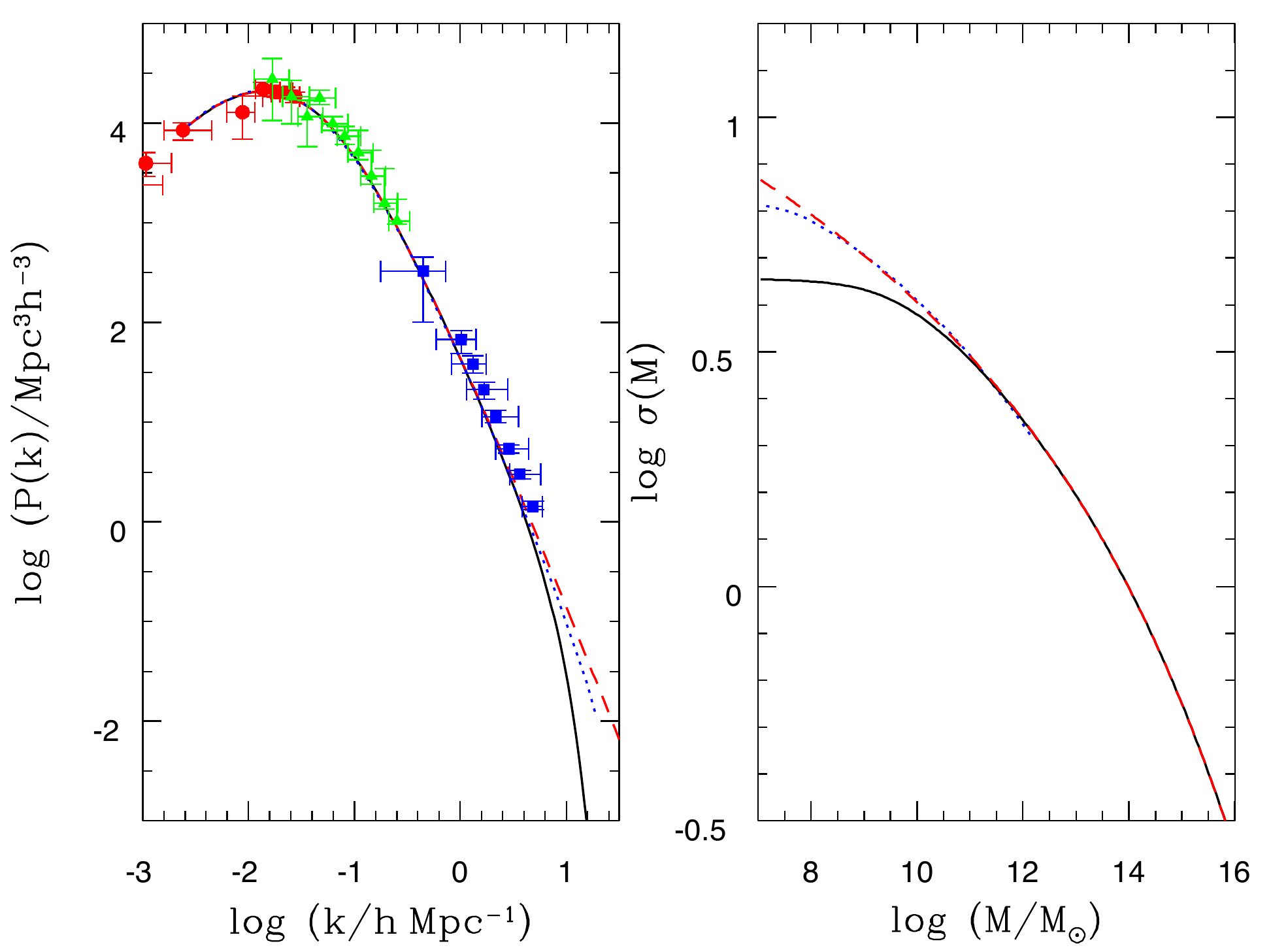}
\caption{Left panel: linear power spectrum computed for 0.75 keV WDM particles (black solid line), for 3.3  keV WDM particles (blue dotted line) and 
for CDM particles (red dashed line). The points with error bars are a compilation of observational data from  Tegmark \& Zaldarriaga 
(2002, and references therein). Right panel: root mean square amplitude of perturbations as a function of the mass scale; the curves correspond to 
the power spectra shown in the left panel. }
\label{fig_ps} 
\end{figure*}

Although  the latest results from the comparison of the observed Ly-$\alpha$ forest with N-body simulations yield $m_X>3.3$ keV  (Viel et al. 2013), thus 
exceeding our adopted value (see the Discussion in Sect. 4), 
our reference value has been chosen to maximize the effect of adopting a WDM spectrum, in order 
to enlighten the role of the power spectrum in the age distribution of stellar populations and on the galactic star formation histories. 
As discussed in recent works (Viel et al. 2013, Schneider et al. 2014), the effects of 
a WDM particle mass consistent with the Lyman-$\alpha$ constraints are nearly indistinguishable from 
CDM, and therefore it is not possible in this case to alleviate any of the the small scale problems. 
This is also visible from Fig.~\ref{fig_ps} where, in the left panel, we compare the power spectra 
calculated for a CDM cosmology (solid line) and for a WDM particle mass 
$m_{X}=3.3$ keV (dotted line) and  $m_{X}=0.75$ keV (dashed line), whereas in the right panel we show the 
amplitude of perturbations as a function of the mass scale calculated in these three cases. The latter quantity is 
calculated as: 
\begin{equation}
\sigma^2(M)= \int \frac{k^2}{2 \pi}\,P(k) \, W(kr) \, dk,
\label{sigma_eq}
\end{equation}
where $M\sim 1.2 \times 10^{12} h^2 M_{\odot} (r/Mpc)^{3}$ is the total mass within a sphere of radius $r$ and $W(kr)$ 
is the top-hat window function (Norman 2001). \\
The merging rates of the SAM are obtained through Monte Carlo realizations, 
where the EPS probabilities depend on the power spectrum only through Eq. ~\ref{sigma_eq}. 

As visible from the right panel of Fig.~\ref{fig_ps}, at the physical scales studied in this work (i.e. at DM halo masses $M>10^8 M_{\odot}$), the rms 
amplitude of perturbations calculated 
assuming a WDM particle with mass $m_X=3.3$ keV 
is indistinguishable from the CDM one hence, from the statistical point of view,  
the merger trees will be nearly identical to those generated in the CDM case. For this reason, 
as far as 
the observable properties of baryons are concerned, in this case we expect results very close 
to those achieved with a CDM cosmology. 

Assuming $m_X=0.75$ keV, we can investigate the effects of the largest suppression of the DM power spectrum. 
This assumption is consistent with the previous lower bound on $m_X$ 
set by the observed Lyman-$\alpha$ forest in absorption spectra of Quasars at $z=2-3$ compared to 
results of N-body simulations  (Viel et al. 2005), indicating $m_X\gtrsim 0.6$ keV for thermal relic particles. 

As visible in Fig.~\ref{fig_ps}, adopting the above value for the WDM particle mass yields a suppression of the power spectrum 
(with respect to CDM) at scales below $r_{fs}\approx 0.2$ Mpc, corresponding to DM haloes with mass $M \lesssim  10^{9.5}\,M_{\odot}$ 
(Menci et al. 2012; 2013). \\
The starting mass grid consists of 20 DM haloes and for each of
them different realizations of the merger trees are generated. 
The galaxy samples obtained in the $\Lambda$CDM and $\Lambda$WDM cosmologies 
consist of 105950 and 20367 objects, respectively. 
The minimum stellar mass resolved in the model is of the order of $10^5 M_{\odot}$. \\
For a more detailed description of 
the detailed procedure adopted to generate Monte Carlo realizations of merger trees within SAMs, 
the reader is referred to Menci et al. (2012).

\subsection{The Baryonic Physics}

The evolution of galaxies is described by 
the collapse and subsequent merging history of the
peaks of the primordial density field, which enables one to generate a
synthetic catalogue of model galaxies and of their past merging
history. 
Galaxies included into larger DM haloes may survive as satellites,
or merge to form larger galaxies due to binary aggregations, or coalesce
into the central dominant galaxy owing to dynamical friction (Menci et al. 2005). 

The evolution of the gas and stellar content is followed for each sub-halo hosting a galaxy. 
The gas which has radiatively cooled, characterised by a mass $m_c$, settles into a rotationally supported disk and condenses 
into stars at a rate $\psi\propto m_c/\tau_d$, where $\tau_d$ is the dynamical time. 
The stellar feedback returns part of the cooled gas to the hot gas phase, i.e. the gas 
at the virial temperature of the halo. At each time step, the mass $\Delta m_h$ returned from the cold
gas content of the disk to the hot gas phase due to Supernovae (SNe) activity is
estimated from standard energy balance arguments (Kauffmann 1996, see also Dekel \& Birnboim 2006) as 
\begin{equation}
\Delta m_h=E_{SN}\,\epsilon_0\,\eta_0\,\Delta m_*/v_c^{2} 
\label{m_h}
\end{equation}
where $\Delta m_*$ is the
mass of stars formed in the timestep, $\eta\approx 7\cdot 10^{-3}/M_{\odot}$
is the number of SNe per unit solar mass for a Salpeter (1955) IMF, 
$E_{SN}=10^{51}\,{\rm ergs}$ is the total energy released by each SN, $v_c$
is the circular velocity of the galactic halo, and $\epsilon_0=0.01$ is the
efficiency of the energy transfer to the cold interstellar gas. The
above mass $\Delta m_h$ is made available for cooling at the next timestep. 
In the present paper, we use the same choice of free parameters as in Menci et al. (2012). \\
In the model, an additional channel for star formation is represented by interaction-triggered
starbursts (ITS), driven by merging or by fly-by events between
galaxies. Such a star-formation mode can convert large fractions (up
to 100 per cent in major merging events) of the cold gas on short
time-scales, of the order of $\sim 10^6$ yr.  This ingredient provides an important contribution to the
early formation of stars in massive galaxies (Menci et al. 2004).

The model also includes a detailed treatment of the growth of
supermassive black holes at the centre of galaxies, of the
corresponding AGN activity powered by the interaction-triggered accretion 
of cold gas onto the black holes, and of the AGN feedback onto the
galactic gas. 
The effects of the AGN on the star formation history of 
galaxies and on their observable properties is discussed 
in Menci et al. (2006); Menci et al. (2008) and Calura \& Menci (2011). 

Galaxy magnitudes are computed by convolving the 
star formation histories (SFHs) with synthetic, metallicity dependent spectral libraries of stellar populations
(Bruzual \& Charlot 2003, hereinafter BC03).

For each galaxy, characterised at the time $t$ by a star formation rate $\psi(t)$, 
the average mass-weighted age at the time $t_0$ can be calculated as 
\begin{equation}
Age\,(t_0) = \frac{\int (t_0-t) \psi(t) dt}{\int \psi(t) dt}.
\label{Age_mw} 
\end{equation} 

If $L^{SSP}_r(t_0-t)$ represents the $r$-band luminosity of a simple stellar 
population of age $(t_0-t)$, the average luminosity-weighted age of the galaxy is: 

\begin{equation}
Age\,(t_0) = \frac{\int (t_0-t)  \psi(t) L^{SSP}_r(t_0-t) dt}{\int \psi(t) L^{SSP}_r(t_0-t)dt}.
\label{Age_mw} 
\end{equation}


\section{Results}
The results presented in this Section 
are computed for two different cosmologies, $\Lambda$-WDM  ($\Lambda$WDM) and $\Lambda$CDM, 
and with the same choice of the parameters related to the baryonic physics. 
This approach is useful to single out the effects of the DM power spectrum on the 
SFHs of present-day galaxies and on a few fundamental observables 
related to their stellar populations. 
The most relevant quantities in our study are the mass-weighted and luminosity-weighted 
integrated ages of the stellar populations. 
The age-mass relation and the light-weighted age distribution calculated by means of our model 
within the two different cosmological scenarios will be compared to data from the largest 
observational sample available, i.e. those from the Sloan Digital Sky Survey. 
We will also study the effects of the different cosmologies on the star formation histories 
of galaxies in various mass bins.

\subsection{The Age-Mass relation}
\label{sec_am}
The integrated ages of the stellar populations of local galaxies are a
measure  of their star formation history and offer a unique way to
assess the typical formation timescales of their basic constituents.
The integrated spectra of local galaxies are commonly used  to
investigate ages by means of population synthesis models  (e.g., 
Bruzual \& Charlot 2003). 
From these studies, degenerate results are usually
derived, owing to the similar effects of age,
metallicity and dust attenuation on the  colours of simple and
composite stellar populations.  
A comprehensive study of this subject has been made 
by Gallazzi et al. (2005).
For 170 000 galaxies of stellar mass $>10^9$ M$_{\odot}$ drawn 
from the Sloan Digital Sky Survey Data Release Two (SDSS DR2) and including galaxies of various morphological types, 
Gallazzi et al. (2005) provided median-likelihood estimates of stellar metallicities, 
light-weighted ages and stellar masses. 
For each parameter, these authors determined the full probability
density function (PDF). 
The median of the PDF represents the fiducial estimate of
the parameter, while the associated uncertainty is quantified by 
the 16th and 84th percentile of the distribution. 
The PDF 
has been derived by comparing the observed strength of spectral
absorption features with those estimated by means 
of the Bruzual \& Charlot (2003) population synthesis models, assuming a Chabrier (2003) initial mass function, convolved with a Monte Carlo library of 150000 SFHs and 
metallicities. The template SFHs are modelled by
an exponentially declining SFR, with varying time of onset and
time-scale, to which random bursts of various intensity and duration
are overlapped. The set of absorption features
used to constrain the PDF includes the 4000 $\AA$ break and the Balmer
lines as age-sensitive indices, as well as [Mg$_2$Fe] and [MgFe]' as
metal-sensitive indices. 

In this work we are interested in the relation between the mean stellar age and the galaxy stellar mass. 
We use the $r$-band luminosity weighted age as a function of stellar mass obtained in Gallazzi et al (2005). 
We complement this here with the relation between the mass-weighted age, as estimated in Gallazzi et al (2008), and the stellar mass. 
For more details and a more complete discussion of systematic effects,
the reader is referred to Gallazzi et al. (2005) and Gallazzi et al. (2008). \\
As found by Gallazzi et al. (2005), the more massive the galaxy, the older 
its average stellar populations: this property of local galaxies  is sometimes refereed to 
as `Archaeological Downsizing' (Fontanot et al. 2009). 

In the model galaxies as well as in the observed sample,  
no selection has been performed on the basis of colour or specific star formation rate. 

\begin{figure*} 
\includegraphics[width=170mm,height=170mm]{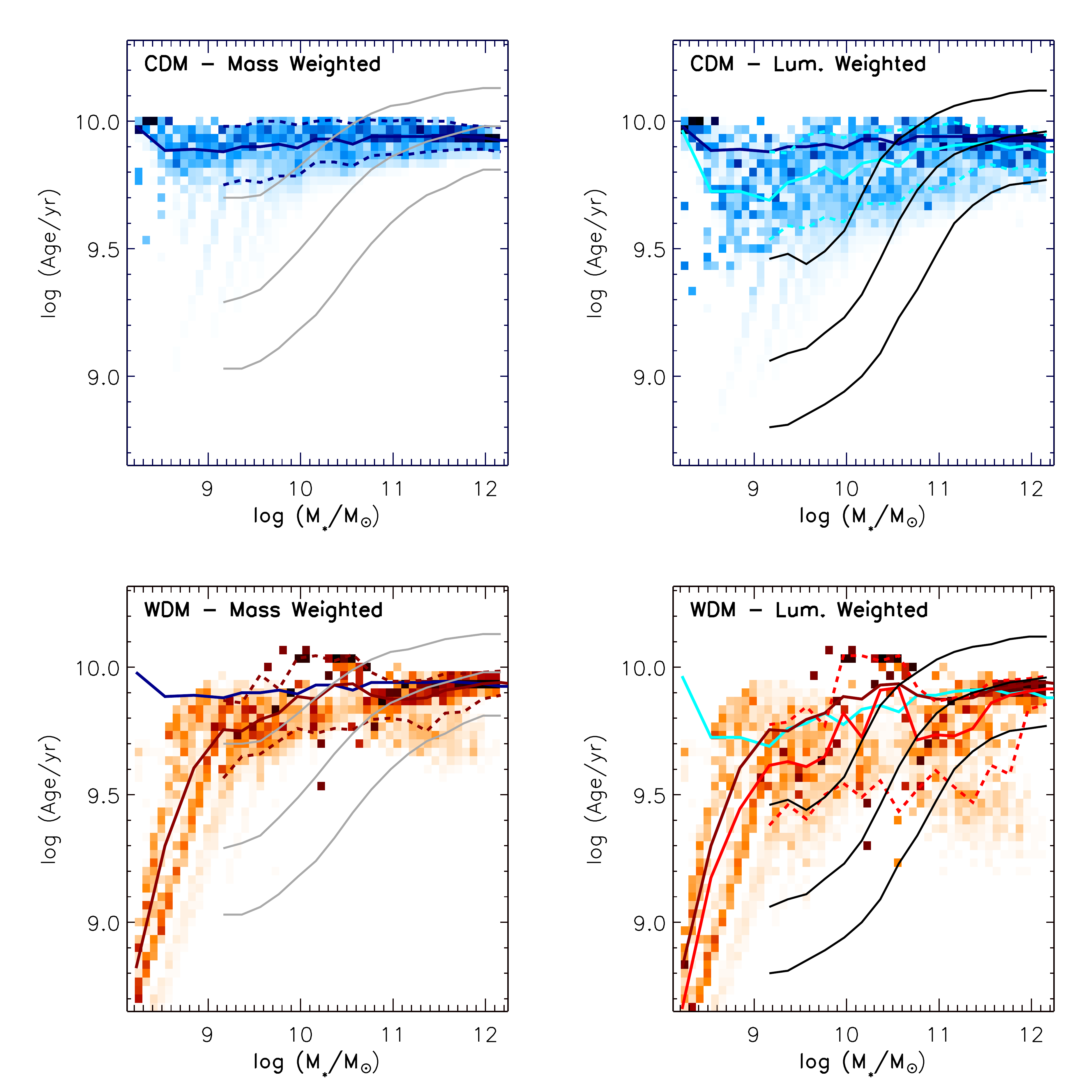}
\caption{The mass-weighted age-mass relation (Left Panels) and the 
$r$-band light weighted age-mass relation (Right Panels) calculated for a  $\Lambda$CDM  (Top Panels) 
and a $\Lambda$WDM (Bottom Panels) cosmology. 
At each stellar mass, the colour-coded regions 
represent the predicted distribution, normalised 
to the total number of galaxies characterised by that stellar mass value. 
The upper, the middle and the lower 
grey (black) curves represent the 16th, the 50th (median) and the 84th percentiles of the 
observed distribution in mass- (light-)weighted stellar age 
(Gallazzi et al. 2008 [2005]), respectively.
The blue, cyan, dark-red and red solid curves represent 
the predicted median, mass-weighted relation in the $\Lambda$CDM case, 
the luminosity-weighted relation in the $\Lambda$CDM case, 
the mass-weighted relation in the $\Lambda$WDM case, 
and the luminosity-weighted relation in the $\Lambda$WDM case, respectively. 
In each panel, within the range of the observed distributions, 
we also plot as dashed lines the P16 and P84 percentiles of the 
predicted distributions, calculated taking into account the observational uncertainties 
(see text for details). }
\label{fig_am} 
\end{figure*}

The observed mass-weighted and light-weighted age-stellar mass relations calculated for the SDSS sample by 
Gallazzi et al. (2005; 2008) are shown in Fig~\ref{fig_am}, together with their dispersion 
and with the results obtained by 
means of our SAM within a $\Lambda$CDM cosmology and a $\Lambda$WDM cosmology.  
The observed age-mass relations and the associated dispersion are 
plotted as 
the 16th, the 50th (median) and the 84th percentiles of the observed distribution in stellar age. 

In order to have a better comparison between 
theoretical results and observations, we have weighted the 
distributions obtained for our model galaxies by the observational uncertainties. 
In each mass bin, the theoretical 
distribution of stellar ages has been convolved with a Gaussian function, 
where the standard deviation is 
represented by the average observational error in that mass bin 
 obtained by Gallazzi et al. (2005). 
We have then computed the cumulative distribution, and 
taken the 50 percentile as the median age in each mass bin. The resulting distributions 
are represented by the coloured solid lines Fig. ~\ref{fig_am} (see caption of Fig. ~\ref{fig_am} for details).  
The dashed lines represent instead the 16$^{th}$ and 84$^{th}$ 
percentiles of the theoretical distributions.

Explaining the slope of the observed  age-mass relation 
has been traditionally challenging within $\Lambda$CDM galaxy formation theories. 
As shown in a few previous works (Fontanot et al. 2009; Pasquali et al. 2010; De Lucia \& Borgani 2012), 
within a $\Lambda$CDM cosmology, the predicted age-mass relation is shallower than the one 
derived observationally, i.e. on average, galaxies of any mass present older ages than those estimated observationally. 
This is confirmed by our results calculated within a $\Lambda$CDM cosmology, as visible in the top-left panel of Fig. ~\ref{fig_am}, 
where we show the present  mass-weighted average age-stellar mass relation compared to the observational data. 

In the top-right panel we compare the theoretical $r$-band luminosity-weighted relation 
computed within a $\Lambda$CDM cosmology to the mass-weighted relation. 
For the most massive galaxies ($M_{*}>10^{11} M_{\odot}$), the predicted ages 
are in good agreement with the median of the 
observational distribution. 
In this kind of studies, obtaining a significant presence of old stellar populations in the most massive galaxies 
is generally not the main difficulty: the most massive galaxies sit in initially higher density peaks, with the natural 
consequences of an early start of star formation and a fast evolution in these systems (Bernardi et al. 2003; De Lucia et al. 2006; Lagos et al. 2009). 
At the massive end, the mass-weighted and luminosity-weighted ages are comparable. 
Again, this is not a new result: 
the small difference is due to the fact that in 
the most massive systems the contribution of young stellar populations to the total 
$r-$band luminosity is not significant, hence the old stellar populations dominate both present mass and luminosity. 

Smaller galaxies present generally larger fractions of young stellar 
populations, which dominate the total $r-$band luminosity. This produces 
lower luminosity-weighted ages in galaxies of smaller masses and a consequently increasing difference between 
the luminosity- and mass-weighted ages with decreasing mass. 
Within a $\Lambda$CDM cosmology, model 
galaxies of $\sim 10^9 M_{\odot}$ are, on average, approximately 4 times older than real galaxies of the same mass. 

In the bottom-left panel of Fig.~\ref{fig_am}, the mass-weighted age-mass relation is shown 
when a $\Lambda$WDM cosmology is considered. The comparison between the same quantity computed within $\Lambda$CDM 
shows already a considerable improvement in terms both of zero-point and slope of the predicted relation. 
Similar results are achieved when the $r$-band luminosity-weighted age-mass relation within $\Lambda$WDM is 
considered (bottom-right panel of Fig.~\ref{fig_am}). In this case, the most massive galaxies present ages 
still in agreement with the observational estimates, although now they appear 
younger than real galaxies according to the mean observational estimates of Gallazzi et al. (2005). 
The most remarkable improvement concerns galaxies of $M_{*}<10^{10.5} M_{\odot}$, for which 
the discrepancy between model results and observations appears reduced: 
in this case, $\sim 10^9 M_{\odot}$ galaxies are $\sim 3  times older$ than their observational analogues. 
This results is encouraging, in particular if one considers 
both the mass-weighted and $r$-band light-weighted age-mass relation at masses $< 10^9 M_{\odot}$: 
in this range, while the $\Lambda$CDM age-mass relation is still shallow, 
within the $\Lambda$WDM cosmology the trend is correct, with decreasing ages at lower masses. \\
To summarize the content of 
this Section, we showed that within a $\Lambda$WDM cosmology 
galaxies follow a steeper age-mass relation than within a $\Lambda$CDM universe, in better agreement with 
the one estimated observationally. 

The impossibility of fully accounting for the observed age-mass relation is 
likely related to our incomplete understanding of the baryon physics. 
Henriques et al. (2013) has shown that by introducing in a fully cosmological, 
$\Lambda$CDM-based SAM 
a new physical mechanism which allows low-mass galaxies to reincorporate the ejected gas over
long timescales, it is possible to have present-day dwarf galaxies 
younger, bluer and more strongly star forming than previously.  
Their study shows also that some small-scale problems tend to persist, 
in particular in the still 
too large fraction of faint, passive dwarf galaxies with respect to observations. 
In principle, the inclusion of similar prescriptions in a WDM-based SAM like the one used here 
could further improve the agreement between the predicted and the observational properties 
of dwarf galaxies, and represents an interesting subject for future work.

\begin{figure*} 
\includegraphics[width=170mm,height=170mm]{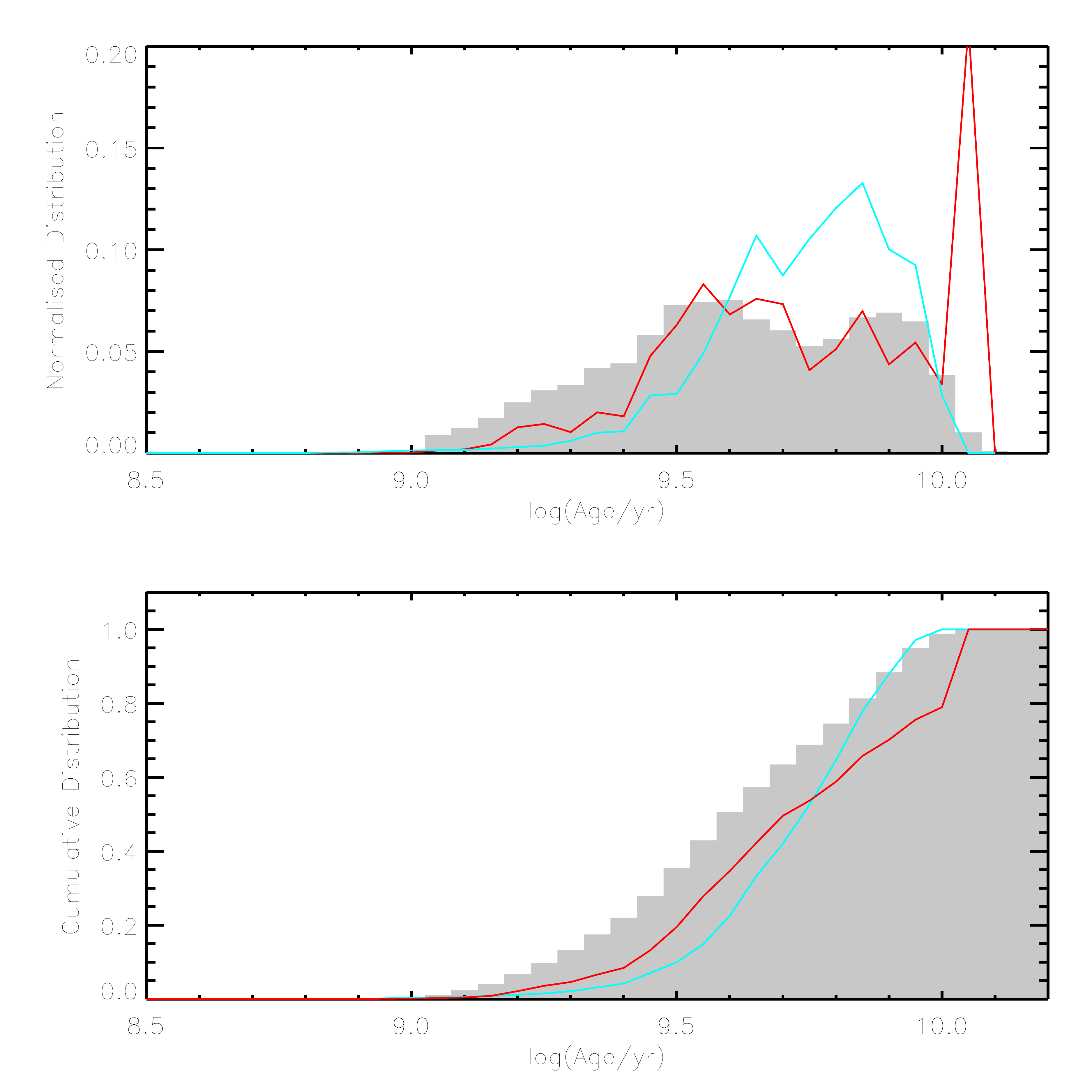}
\caption{Top panel: observed distribution in $r$-band light-weighted age for the 
SDSS sample of Gallazzi et al. (2008, gray shaded histogram) and as predicted by means of 
our semi-analytic model assuming a $\Lambda$WDM cosmology (red curve) and a $\Lambda$CDM cosmology (cyan curve). 
Bottom panel: cumulative age distributions, with all the curves defined as in the top panel.   
Only galaxies more massive than $10^9 M_{\odot}$ are included in this plot.}
\label{fig_distr}
\end{figure*}

\subsection{The Age distribution}
\label{sec_aged}
In this section we discuss the distribution in luminosity-weighted age of 
the entire population of local galaxies more massive than $10^9 M_{\odot}$, as obtained in 
Gallazzi et al (2008). 
The sample of Gallazzi et al. (2008) represents an extension of the 
sample of Gallazzi et al. (2005), which includes only galaxies with signal-to-noise (S/N) ratios greater 
than 20 and which therefore is biased towards more concentrated, high-surface brightness objects. 
In order to study stellar metallicities and ages also in 
low-surface brightness galaxies, in the study of Gallazzi et al. (2008) 
the individual spectra of galaxies with S/N$\le 20$ have been co-added, improving the quality 
of the data for low-surface brightness galaxies. 
While including these galaxies has little effect on the age distribution at fixed mass, it is necessary in order 
to derive a complete census of the ages of local galaxies.
To calculate the distributions shown in this paper, each galaxy spectrum has been weighted by 1/V$_{max}$, where 
V$_{max}$ is defined as 
the maximum visibility volume given by the bright and
faint magnitude limits of the observational sample, $14.5\le r \le17.77$. 

In the top panel of Figure~\ref{fig_distr} we show the total observed 
$r-$band light-weighted 
stellar age distribution as derived by Gallazzi et al. (2008), 
compared to the theoretical distributions calculated 
within a $\Lambda$CDM and a $\Lambda$WDM cosmology. 
All the age distributions plotted in the top panel of figure~\ref{fig_distr} 
are normalised so that the total area under each curve is equal to one. 

The luminosity-weighted ages of real galaxies 
follow an asymmetric, bimodal distribution showing one broad 
peak at age 4 Gyr, accompanied by a slightly lower peak centered at $\sim 8$ Gyr. 

With a $\Lambda$CDM cosmology, 
the age distribution shows two very close peaks 
centered at $\sim 4.5$ and $\sim 7$ Gyr. 
The theoretical distribution is narrower than the observational one, 
with the largest discrepancies between the two visible at the lowest ages (i.e. at ages $\le 3$ Gyr). 

The age distribution obtained for $\Lambda$WDM galaxies shows a broad peak centered at $\sim 4$ and a 
much  higher, narrower peak centered at $\sim 10$ Gyr.  
This narrow peak is due to the presence of a very 
small number of galaxies presenting strong star formation rates 
at early times. Their signature is visible also in the bottom-right panel of Fig.~\ref{fig_am} 
as a relative density maximum in the mass range $9.5 \lesssim log(M_{*}/M_{\odot}) \lesssim 10.5 $ and at ages 
$10^{10}$ yr. 
Further details on these galaxies will be discussed later in Sect.~\ref{sec_sfh}, when we will 
analyze the star formation histories of our model galaxies. 

The same galaxies belong to the lowest-mass dark matter haloes of the $\Lambda$WDM sample, 
and they are characterised by a large cosmological weight, 
thus leaving visible signs also in the computed age distribution. 

The age distribution of $\Lambda$WDM shows a more pronounced tail at low ages with respect to $\Lambda$CDM galaxies. 
For ages less than $\sim 5$ Gyr, 
the overall shape of the observed age distribution is better captured in the $\Lambda$WDM case, as 
visible also from the lower panel of figure~\ref{fig_distr}, where the cumulative age distributions are shown. 
At larger ages, the discrepancy between the observed cumulative distribution and the one predicted in the $\Lambda$WDM is 
due to the already mentioned few peculiar galaxies of the $\Lambda$WDM sample. 
Despite this discrepancy, the similar general aspect of the predicted age distribution of 
galaxies within a $\Lambda$WDM cosmology 
and the one found for real galaxies can be regarded 
as a further encouraging result which corroborates our analysis. 
The conclusion of this analysis is that, by adopting a $\Lambda$WDM cosmology,  
significant steps forward can be made in accounting for the average ages of local galaxies, 
a task traditionally difficult within a $\Lambda$DCM cosmology.

\subsection{The star formation histories}
\label{sec_sfh}
\begin{figure*}
\centering
\includegraphics[width=150mm]{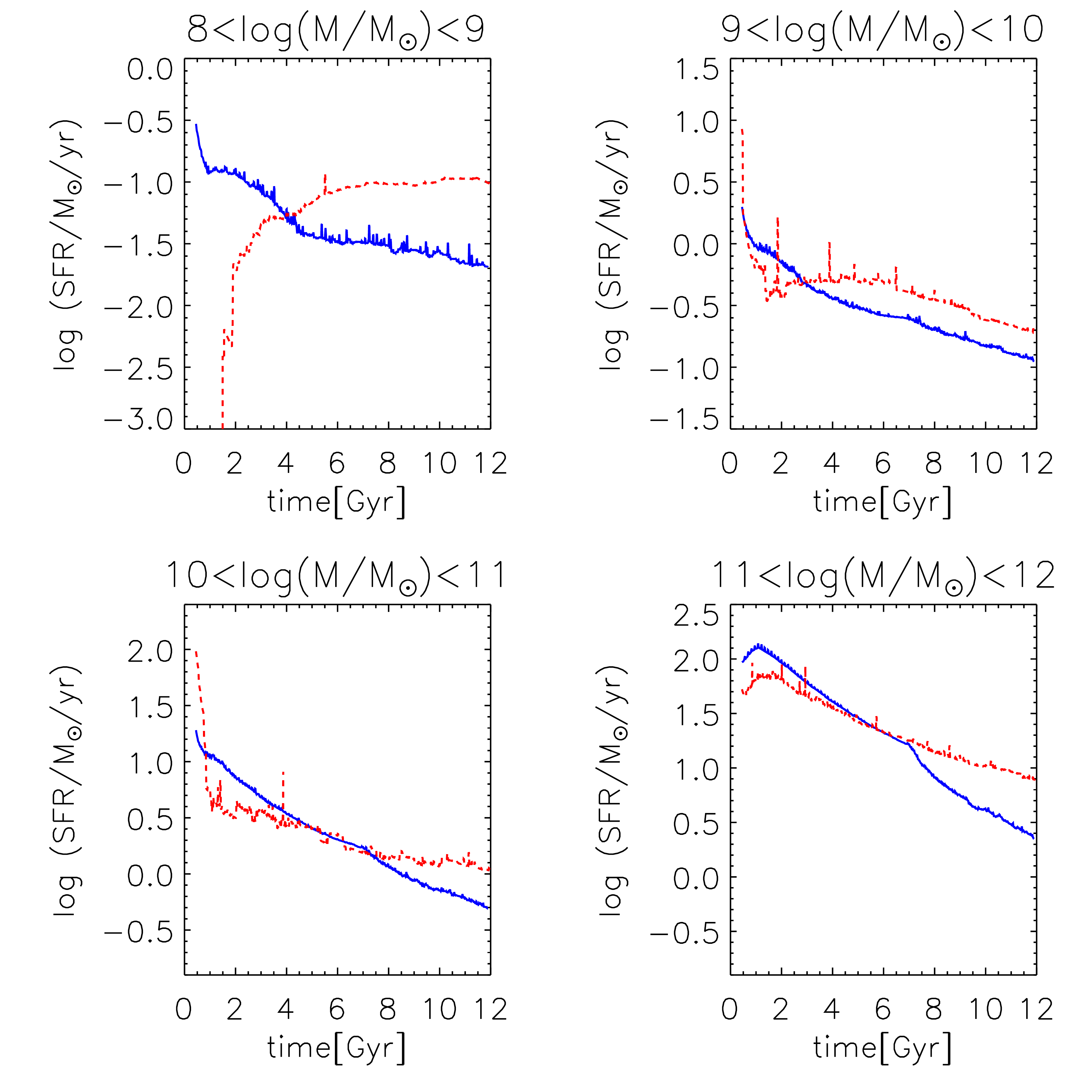}
\caption{Average star formation histories of our simulated galaxies divided 
into four stellar mass bins (reported on  top of each panel), 
computed adopting a $\Lambda$WDM cosmology (dashed, red curves) and a $\Lambda$CDM cosmology (blue solid curves).}
\label{fig_sfh}
\end{figure*}

In the previous sections, we have shown the effects of the underlying 
cosmological framework, in particular of the type of dark matter, 
on the ages of stellar populations in local galaxies. 
In this Section, we aim at studying how the 
underlying cosmology affects the star formation histories. 
We divide our simulated galaxies into four mass bins, and in each bin, 
we compute the star formation rate, at
any given time, averaged over all the galaxies included in that bin, 
taking into account the single SFH of each progenitor. 
The average star formation histories computed in the mass bins are 
shown in Fig.~\ref{fig_sfh}. 

The adoption of a $\Lambda$WDM cosmology affects mostly low-mass galaxies: 
in the lowest mass bin, star formation appears globally delayed with respect to the $\Lambda$CDM. 
In this mass bin, the underlying cosmology affects the overall shape of the average star formation history: 
while in the $\Lambda$CDM the SFH peaks at the earliest 
times and then decreases progressively, in the $\Lambda$WDM case the behaviour is reversed, in that the 
average SFH increases with cosmic time. 
This could be expected from the results discussed in Sect.~\ref{sec_am}, in particular from the 
strong effects of the adopted cosmology on the ages of the stellar population in low-mass galaxies. 

No significant effect is visible in the average star formation histories of the intermediate mass bins  
($9<log(M/M_{\odot})<10$ and $10<log(M/M_{\odot})<11$ ), where the only appreciable differences are a narrow, 
short early peak and an overall slightly flatter SFH in the $\Lambda$WDM case. 
The sharp peak visible in the $\Lambda$WDM star formation histories 
is again due to the few 
early-star forming galaxies mentioned in Sect.~\ref{sec_aged}. 
The fact that they are characterised by larger SFR values than their CDM analogues can be 
understood as follows. 
Within a WDM universe, the first baryonic structures form later 
than their CDM counterparts, moreover they are characterised by a larger mass. 
When this happens, their entire gas reservoir is available for star formation, 
giving place to intense starbursts.  
On the other hand, within LCDM, the first 
baryonic structures to collapse are smaller, hence they have less gas available 
for star formation and are therefore characterised by lower SFR values. 
As the free streaming length determines the 
scale below which the perturbations in the primordial density field are wiped out, 
within a WDM universe the first halos to collapse will be those above a certain scale, 
which in this case corresponds to halos with present-day stellar masses of $\sim 10^{9} M_{\odot}$.

The difference in the slope of the star formation histories is more pronounced in the largest mass bin. The flatter average SFH of 
$\Lambda$WDM galaxies implies a slower average gas consumption timescale, accompanied  
by a lower peak at early times and larger SFR values at late times than those characterising $\Lambda$CDM galaxies. 

Younger stellar populations in the most massive galaxies in the $\Lambda$WDM case 
were also found in the age-mass relation discussed in Sect.~\ref{sec_am}. 
Possible mechanisms to decrease the gas-consumption
timescale in large galaxies will need to be studied in the near
future, and we postpone a detailed treatment of this subject to a forthcoming paper. 

\section{Discussion and Conclusions}

At galactic and sub-galactic scales ($\lesssim 1 Mpc$), 
computations based on N-body simulations pose several problems to the $\Lambda$CDM scenario. 
The first problem concerns the number of satellites present in the Milky Way halo, 
which in  $\Lambda$CDM simulations is expected to contain 
between $\sim 5$  and $\sim 10$  more satellites than what current observations indicate.

On the observational side, the Missing Satellite problem could also be 
related to the limited 
capability of modern telescopes to detect the lowest mass galaxies: 
the number of detected Mily-Way satellites is increasing all the time as we
probe fainter objects, and 5-20 times more could be discovered in
the future (Bullock et al. 2010). 
On the other hand, a theoretical explanation for the Missing Satellite problem could also be found 
within the $\Lambda$CDM framework. Wang et al. (2012) have shown that the overabundance of local satellites 
could be related to our poor knowledge of the virial mass of the Milky Way halo,  
and that with a dark matter halo of mass $M_h < 10^{12} M_{\odot}$, the discrepancy 
between simulation results and observations can be reconciled. Sawala et al. (2013) have shown instead that 
sub-galactic baryonic processes could alleviate this tension between simulation results, 
as could large-scale effects such as 'cosmic stripping' (Benitez-Llambay et al. 2013).
There are however other problems which seem more difficult to solve within $\Lambda$CDM. 
Boylan-Kolchin, Bullock \& Kaplinghat (2012) showed that 
the bright satellites of the MW are hosted by 
DM sub-halos significantly less dense than the most massive sub-halos in state-of-the-art 
$\Lambda$CDM simulations. (Boylan-Kolchin, Bullock \& Kaplinghat 2012). 
In this case, even strong baryonic feedback does not seem to 
alleviate this discrepancy.  
The origin of the above problems could be related to 
an excess of power of the CDM power spectrum at small scales; in this case, 
a natural solution is represented by the adoption of a universe dominated by warm dark matter. 

In a previous paper, Menci et al. (2012) have tested for the first time the effects of 
$\Lambda$WDM on the statistical properties of galaxies at various redshifts. These authors 
have shown that 
the adoption of a power spectrum  which is suppressed (compared to the
standard CDM case) below a cut-off scale of $\sim $ 1 Mpc, helps limit the
wealth of substructure typical of LCDM models. 

By means of the same model described in Menci et al. (2012),  
in this paper we have studied the effects of a $\Lambda$WDM power 
spectrum on the ages of the stellar populations and on 
the star formation histories of present-day galaxies. 
We have focused on the age-mass relation, a well-established 
property of local galaxies which outlines one fundamental aspect of their `downsizing' 
nature, i.e. the fact that the largest galaxies host the oldest stellar populations (Gallazzi et al. 2005; Fontanot et al. 2009; 
Pasquali et al. 2010). 
Explaining this relation within `concordance' $\Lambda$CDM cosmological models 
has represented a long-standing problem, for the main reason that 
in the standard cosmological scenario, local low-mass galaxies appear too passive, characterised by 
too low present star formation rates and by an excess of old stars, with respect to 
the observations. 
This manifests in a predicted age-mass relation which is too shallow with respect to the observed one 
(Fontanot et al. 2009; De Lucia \& Borgani 2012), with a reasonable agreement between models and data 
at the bright end, but accompanied by a severe overestimate of the stellar ages of galaxies at masses $M_{*}<10^{10} M_{\odot}$. 
This discrepancy is partly related to the well-known difficulties of modelling baryonic processes in cosmological 
models: star formation, gas cooling and energetic feedback from exploding supernovae are highly 
non-linear processes and our present understanding of each of them is far from being complete. 
In this regard, a step forward is represent by the study of Henriques et al. (2013) who showed how, 
by means of a new physical mechanisms allowing to reincorporate the gas ejected in the lowest mass galaxies, 
it was possible to obtain local dwarf galaxies bluer and more actively star forming than in previous studies. 

However, the inability of $\Lambda$CDM models to describe realistically the star formation histories 
of low-mass galaxies could be due to the underlying $\Lambda$CDM cosmological framework, in which
the smallest structures are the earliest to emerge and, without any realistic physical mechanism  
which delays the baryon cooling and the subsequent star formation, these objects form too large a fraction of their stars 
(compared to the observational estimate) at high redshift (Fontanot et al. 2009). 
In this regard, the choice of a WDM power spectrum may help alleviating several discrepancies 
between the observed properties of galaxies and the model results since, beside implying the 
suppression of the small scale power, it also causes an overall delay of 
the growth of structure formation (Amendola et al. 2013; Destri et al. 2013) and of star formation. 
Note however that, in order to enlighten the impact of a  WDM power spectrum on the age distributions, 
we have adopted in this paper a particle mass corresponding (for thermal relics) to $m_X=0.75$ keV.
Several observational constraints indicate larger values for the dark matter particle mass: 
from long GRBs at high redshifts, de Souza et al. 2013 
suggest $m_X>1.6$ keV at 95 per cent confidence level. 
From the analysis of image fluxes in gravitationally lensed quasi-stellar objects,  
Miranda \& Macci{\`o} (2007) find a lower limit of 10 keV for the mass of WDM candidates in the form of sterile neutrinos. 
More constraints come from various recent studies based on the extensive use of N-body simulations. 
From the distribution of satellites in the MW, Macci{\`o} \& Fontanot (2010) suggest a lower limit of 1 keV for thermal WDM particles. 
The latest constraints from the observed absorption in 
the Lyman-$\alpha$ forest indicate a lower limit of $m_X\gtrsim 3.3$ keV for thermal relic particles (Viel et al. 2013; Markovic \& Viel 2013).
From high-redshift galaxy counts, Schultz et al. (2014) suggest that 
a 1.3 keV WDM particle can be ruled out at high significance.  
Overall, most of these results would indeed 
leave very little room for a contribution 
of WDM to the solution of the small scale crisis of CDM (see, e.g., Polisensky \& Ricotti 2013; Herpich et al. 2013). 
However, it is worth stressing that artificial fragmentation is a serious 
issue in current N-body numerical simulations on scales around and below the free streaming length
(Lovell et al. 2012, Schneider et al. 2013, Hahn et al. 2013; Markovic \& Viel 2013; Lovell et al. 2013), and that  
increasing the resolution even by rather large factors is not sufficient to prevent 
this effect (Wang \& White 2007). 
The recent study of Lovell et al. (2013) uses an algorithm to identify spurious clumps 
and to remove them from the catalogues and, from the constrains provided by the abundance of 
Milky-Way satellites, they obtain a conservative lower limit to the WDM thermal particle mass $m_X$=1.5 keV. 
The use of a similar algorithm also in other studies based on N-body simulations will be 
crucial to investigate further the systematic effects of artificial fragmentation, and to eventually have 
more robust constraints on the WDM particle mass candidate. 

In such a context, the scope of the present paper is not to provide a test for the WDM scenario, nor to pin down the 
DM candidate that provides the power spectrum that best matches the observations. Rather, we aim at showing that 
{\it assuming} that the solution of the CDM small scale crisis relies in a cut-off in the power spectrum, such a  
cut-off {\it also} provides a viable solution to the problems related to the age distribution of stellar populations. 
In particular, the assumption of a DM power-spectrum cutoff such as that provided by a WDM particle with mass $m_X\approx 1$ keV 
has a significant impact on the age distribution of stellar populations in local galaxies as detailed below:

\begin{itemize}
\item Within a $\Lambda$WDM cosmology, both the mass-weighted and the 
luminosity-weighted age-mass relation are steeper than those obtained within a $\Lambda$CDM universe, in better agreement with 
the one estimated observationally. 
As expected, $\Lambda$WDM galaxies of all masses  present, on average,  younger stellar populations 
than their $\Lambda$CDM counterparts. At the massive end, $\Lambda$WDM galaxies appear  
younger than real galaxies. 
\item Our study of the average age distribution of local galaxies 
has confirmed that the best results can be obtained within a LWDM cosmology, assuming a particle mass of 0.75 keV: 
in this case, the bimodal behaviour of the observational distribution is reproduced, 
although differences in the width and in the peak ages between the predicted distribution and the 
observed one are visible. 
\item We studied the effects of a WDM power spectrum on the 
average star formation histories of the simulated galaxies. 
We grouped the present-day stellar masses of the model galaxies into four 
different mass bins and in each bin, we computed an average star 
formation rate. The adoption of a $\Lambda$WDM cosmology affects mostly low-mass galaxies, for which 
we find a star formation history significantly different with respect to the $\Lambda$CDM. 
While in the $\Lambda$CDM the SFH peaks at the earliest 
times and then decreases progressively, in the $\Lambda$WDM case the behaviour is reversed, in that the 
average SFH increases with cosmic time. 
This result seems in qualitative agreement with recent attempts to reconstruct the 
SFHs of blue galaxies by means of physically motivated, comprehensive libraries of 
synthetic spectra (Pacifici et al. 2013). 
Significant differences are visible also in the highest mass bin ($11<log(M/M_{\odot})$), 
where galaxies within $\Lambda$WDM show a lower early peak and larger present-day SFRs than their $\Lambda$CDM counterparts. 
This aspect will need to be improved in the future, by testing possible mechanisms to 
quicken the gas consumption timescales in the most massive systems. 
\end{itemize} 
Other studies carried out by means of semi-analytical models have demonstrated the importance of
the choice of the parameters related to baryonic physics in determining some 
fundamental scaling relations and properties of galaxies (Kang et al. 2013) and AGNs (Menci et al. 2013) 
within WDM models.  
Kang et al. (2013) outlined the difficulty in disentangling between 
the true effects due to WDM and the specific implementation of baryonic physics when 
observables such as luminosity functions, the ratio between  between DM and stellar mass in galaxies and, most of all, 
dynamical mass estimators such as the Tully-Fisher relation are considered. 
Menci et al. (2013) studied the evolution of the AGN LF and showed how 
a $\Lambda$CDM model with extreme assumptions for both AGN and stellar feedback can produce results very similar to 
those obtained by means of a WDM, in particular at low redshift. 
The tests performed in this paper have shown that the stellar populations in 
galaxies of stellar mass $<10^{10}$ M$_{\odot}$, characterised by a relatively simpler physics 
owing mostly to the minor effects of AGNs on their star formation histories, 
may help disentangling between the effects of a WDM power spectrum and the baryonic physics. 
We have shown that a WDM cosmology can naturally account for 
younger stellar populations in low mass galaxies with respect to 
a $\Lambda$CDM model. 
At present, figuring 
realistic mechanisms which can further delay star formation in low-mass systems at high redshift 
seems difficult and may require substantial revisions in the current feedback schemes 
(Fontanot et al. 2009, Weinmann et al. 2012, Henriques et al. 2013).

\section*{Acknowledgments}
We wish to thank Carlo Nipoti for useful comments on an early version of the paper, Lauro Moscardini 
for interesting discussions 
and an anonymous referee for several comments that improved the paper. 
FC acknowledges financial support from PRIN MIUR 2010-2011, 
project "The Chemical and Dynamical Evolution of the Milky Way and Local Group Galaxies", 
prot. 2010LY5N2T. 
AG acknowledges support from the European Union Seventh Framework Programme (FP7/2007-2013) under grant
agreement n. 267251.

%
%

\label{lastpage}

\end{document}